\documentclass[12pt,a4paper]{article}
\usepackage{amsfonts}
\usepackage{amsmath,amssymb,amsthm} 
\usepackage{graphics}
\usepackage{dsfont}
\usepackage{float}
\usepackage{url}
\usepackage{color}
\usepackage[table]{xcolor}
\usepackage{epsfig}
\usepackage{natbib}
\usepackage[top=1in, left=1in, right=1in, bottom=1in]{geometry}
\usepackage{setspace}

\bibpunct[, ]{(}{)}{,}{a}{}{,}

\theoremstyle{plain}
\newtheorem{theo}{Theorem}

\theoremstyle{definition}

\theoremstyle{remark}

\def\1{\mathds{1}}

\renewcommand{\leq}{\leqslant}

\renewcommand{\geq}{\geqslant}

\input{tcilatex}
\begin{document}

\title{\textbf{Quantile of a Mixture} }
\author{Carole Bernard\thanks{%
Carole Bernard, Department of Statistics and Actuarial Science at the
University of Waterloo (email: \texttt{c3bernar@uwaterloo.ca}). }\ \ and
Steven Vanduffel\thanks{\underline{Corresponding author :} Steven Vanduffel,
Department of Economics and Political Sciences at Vrije Universiteit Brussel
(VUB). (e-mail: \texttt{steven.vanduffel@vub.ac.be}). } \thanks{
C. Bernard gratefully acknowledges support from the Natural Sciences and
Engineering Research Council of Canada, the Humboldt Research Foundation and
the hospitality of the chair of mathematical statistics of Technische
Universit\"{a}t M\"{u}nchen where the paper was completed. S. Vanduffel
acknowledges the financial support of the BNP Paribas Fortis Chair in
Banking. }}
\date{ \today }
\maketitle

\begin{abstract}
In this note, we give an explicit expression for the quantile of a mixture
of two random variables. We carefully examine all possible cases of discrete
and continuous variables with possibly unbounded support. The result is
useful for finding bounds on the Value-at-Risk of risky portfolios when only
partial information is available (\cite{BVnew}).
\end{abstract}

\newpage 

\begin{center}
\textbf{\LARGE Quantile of a Mixture} 
\end{center}

Let $X$ and $Y$ be two random variables. We denote by $F_{X}$ and $F_{Y}$
their respective marginal distributions. For $p\in (0,1),$ the quantile of $X
$ at level $p$ is defined as 
\begin{equation}
F_{X}^{-1}(p)=\inf \left\{ x\in \mathbb{R}\;|\;F_{X}(x)\geq p\right\} .
\label{VAR}
\end{equation}%
By convention, $\inf \{\emptyset \}=\infty $ and $\inf \{\mathbb{R}%
\}=-\infty ,$ so that the quantile is properly defined by \eqref{VAR} for
all $p\in \lbrack 0,1]$. In risk management one often considers $X$ as a
loss variable in which case $F_{X}^{-1}(p)$ can be broadly interpreted as
the maximum loss (\textquotedblleft Value-at-Risk\textquotedblright ) one
can observe with $p-$confidence. Value-at-Risk computations are at the core
of setting capital requirements for banks and insurance companies.

Consider a sum $S=\mathbb{I}X+$ $(1-\mathbb{I})Y,$ where $\mathbb{I}$ is a
Bernoulli distributed random variable with parameter $q$ and where the
components $X$ and $Y$ are independent of $\mathbb{I}$. Our objective is to
find an explicit expression for the quantiles of $S$ as a function of the
quantiles of its components $X$ and $Y$. A direct application is to find
bounds on Value-at-Risk in the case when partial information is available (%
\cite{BVnew}).

\begin{theo}[Quantile of a mixture]
\label{lem2new} Consider a sum $S=\mathbb{I}X+$ $(1-\mathbb{I})Y,$ where $%
\mathbb{I}$ is a Bernoulli distributed random variable with parameter $q$
and where the components $X$ and $Y$ are independent of $\mathbb{I}$. Define 
$\alpha _{\ast }\in \lbrack 0,1]$ by 
\begin{equation*}
\alpha _{\ast }:=\inf \left\{ \alpha \in (0,1)\text{ }|\text{ }\exists \beta
\in (0,1)\ \Big\{%
\begin{array}{l}
q\alpha +(1-q)\beta =p \\ 
F_{X}^{-1}(\alpha)\geq F_{Y}^{-1}(\beta) \\ 
\end{array}%
\right\}
\end{equation*}%
and let $\beta _{\ast }=\frac{p-q\alpha _{\ast }}{1-q}\in[0,1].$ Then, for $%
p\in \left( 0,1\right) ,$ 
\begin{equation}
s_{p}:=F_S^{-1}(p)=\max \left\{
F_{X}^{-1}(\alpha_\ast),F_{Y}^{-1}(\beta_\ast)\right\}
\end{equation}
This maximum can be computed explicitly by distinguishing along the four
following cases for $F(\cdot )$ and for $G(\cdot ):$

\noindent {Case 1: $F$ is continuous in $s_{p}$ and for all $z<s_{p}$, $%
F(z)<F(s_{p})$}\newline
\noindent {Case 2: $F$ is continuous in $s_{p}$ and there exists $z<s_{p}$, $%
F(z)=F(s_{p})$}\newline
\noindent {Case 3: $F$ is discontinuous in $s_{p}$ and for all $z<s_{p}$, $%
F(z)<F(s_{p}^{-})$}\newline
\noindent {Case 4: $F$ is discontinuous in $s_{p}$ and there exists $z<s_{p}$%
, $F(z)=F(s_{p}^{-})$}\newline
\noindent {Case a: $G$ is continuous in $s_{p}$ and for all $z<s_{p}$, $%
G(z)<G(s_{p})$}\newline
\noindent {Case b: $G$ is continuous in $s_{p}$ and there exists $z<s_{p}$, $%
G(z)=G(s_{p})$}\newline
\noindent {Case c: $G$ is discontinuous in $s_{p}$ and for all $z<s_{p}$, $%
G(z)<G(s_{p}^{-})$}\newline
\noindent {Case d: $G$ is discontinuous in $s_{p}$ and there exists $z<s_{p}$%
, $G(z)=G(s_{p}^{-})$}

We have summarized the computations of $s_{p}$ in Table \ref{SUM} for the
sixteen possible combinations.

\begin{table}[!htb]
\begin{tabular}{c|c|c|c|c}
| & (a) & (b) & (c) & (d) \\ \hline
(1) & $%
\begin{array}{l}
\alpha_*=F(s_p) \\ 
\beta_*=G(s_p) \\ 
s_p=F_X^{-1}(\alpha_*) \\ 
{\color{white} s_p}=F_Y^{-1}(\beta_*)%
\end{array}%
$ & $%
\begin{array}{l}
\alpha_*=F(s_p) \\ 
\beta_*=G(s_p) \\ 
s_p=F_X^{-1}(\alpha_*) \\ 
{\color{white} s_p}>F_Y^{-1}(\beta_*)%
\end{array}%
$ & $%
\begin{array}{l}
\alpha_*=F(s_p) \\ 
s_p=F_X^{-1}(\alpha_*) \\ 
{\color{white} s_p}=F_Y^{-1}(\beta_*)%
\end{array}%
$ & $%
\begin{array}{l}
\alpha_*=F(s_p) \\ 
\hbox{if}\ F_S(s_p^-)<p, \\ 
\hspace{5mm}s_p=F_X^{-1}(\alpha_*) \\ 
{\color{white} \hspace{5mm}s_p}=F_Y^{-1}(\beta_*) \\ 
\hbox{if}\ F_S(s_p^-)=p, \\ 
\hspace{5mm}s_p=F_X^{-1}(\alpha_*) \\ 
{\color{white} \hspace{5mm}s_p}>F_Y^{-1}(\beta_*)%
\end{array}%
$ \\ \hline
(2) & $%
\begin{array}{l}
\alpha_*=F(s_p) \\ 
\beta_*=G(s_p) \\ 
s_p=F_Y^{-1}(\beta_*) \\ 
{\color{white} s_p}>F_X^{-1}(\alpha_*)%
\end{array}%
$ & Impossible & $%
\begin{array}{l}
\alpha_*=F(s_p) \\ 
s_p=F_Y^{-1}(\beta_*) \\ 
{\color{white} s_p}>F_X^{-1}(\alpha_*)%
\end{array}%
$ & $%
\begin{array}{l}
\alpha_*=F(s_p) \\ 
s_p=F_Y^{-1}(\beta_*) \\ 
{\color{white} s_p}>F_X^{-1}(\alpha_*)%
\end{array}%
$ \\ \hline
(3) & $%
\begin{array}{l}
\beta_*=G(s_p) \\ 
s_p=F_X^{-1}(\alpha_*) \\ 
{\color{white} s_p}=F_Y^{-1}(\beta_*)%
\end{array}%
$ & $%
\begin{array}{l}
\beta_*=G(s_p) \\ 
s_p=F_X^{-1}(\alpha_*) \\ 
{\color{white} s_p}>F_Y^{-1}(\beta_*)%
\end{array}%
$ & $%
\begin{array}{l}
s_p=F_X^{-1}(\alpha_*) \\ 
{\color{white} s_p}= F_Y^{-1}(\beta_*)%
\end{array}%
$ & $%
\begin{array}{l}
\hbox{if}\ F_S(s_p^-)<p, \\ 
\hspace{5mm}s_p=F_X^{-1}(\alpha_*) \\ 
{\color{white} \hspace{5mm}s_p}=F_Y^{-1}(\beta_*) \\ 
\hbox{if}\ F_S(s_p^-)=p, \\ 
\hspace{5mm}s_p=F_X^{-1}(\alpha_*) \\ 
{\color{white} \hspace{5mm}s_p}>F_Y^{-1}(\beta_*)%
\end{array}%
$ \\ \hline
(4) & $%
\begin{array}{l}
\beta_*=G(s_p) \\ 
\hbox{if}\ F_S(s_p^-)<p, \\ 
\hspace{5mm}s_p=F_Y^{-1}(\beta_*) \\ 
{\color{white} \hspace{5mm}s_p}=F_X^{-1}(\alpha_*) \\ 
\hbox{if}\ F_S(s_p^-)=p, \\ 
\hspace{5mm}s_p=F_Y^{-1}(\beta_*) \\ 
{\color{white} \hspace{5mm}s_p}>F_X^{-1}(\alpha_*)%
\end{array}%
$ & $%
\begin{array}{l}
\beta_*=G(s_p) \\ 
s_p=F_X^{-1}(\alpha_*) \\ 
{\color{white} s_p}>F_Y^{-1}(\beta_*)%
\end{array}%
$ & $%
\begin{array}{l}
\hbox{if}\ F_S(s_p^-)<p, \\ 
\hspace{5mm}s_p=F_X^{-1}(\alpha_*) \\ 
{\color{white} \hspace{5mm}s_p}=F_Y^{-1}(\beta_*) \\ 
\hbox{if}\ F_S(s_p^-)=p, \\ 
\hspace{5mm}s_p=F_Y^{-1}(\beta_*) \\ 
{\color{white} \hspace{5mm}s_p}>F_X^{-1}(\alpha_*)%
\end{array}%
$ & Impossible \\ \hline\hline
\end{tabular}%
\caption{Summary of all cases for the quantiles of a mixture where $%
s_p=F_S^{-1}(p)$. In all cases, $\protect\alpha_*$ is defined as 
\eqref{defal} and $\protect\beta_*=\frac{p-q \protect\alpha_*}{1-q}\leqslant
G(s_p)$, $\protect\alpha*=\frac{p-(1-q)\protect\beta_*}{q}\geqslant F(s_p)$.}
\label{SUM}
\end{table}
\end{theo}

\newpage \proof Denote by $F(x)$ and $G(x)$ the distributions of $X$ resp. $%
Y.$ Since $X$ and $Y$ are independent of $\mathbb{I}$ we find for the
distribution of $S=\mathbb{I}X+$ $(1-\mathbb{I})Y,$%
\begin{equation*}
F_{S}(x)=qF(x)+(1-q)G(x)\quad \quad x\in 
\mathbb{R}
.
\end{equation*}%
Let $p\in \left( 0,1\right) $ and denote $F_S^{-1}(p) $ by $s_{p}$,%
\begin{equation*}
s_{p}=\inf \left\{ x\in \mathbb{R}\;|\;qF(x)+(1-q)G(x)\geq p\right\} .
\end{equation*}%
In what follows, when considering $\alpha ,\beta \in (0,1)$ we always assume
that they satisfy $q\alpha +(1-q)\beta =p.$ Note that we define $\alpha
_{\ast }$ as 
\begin{equation}
\alpha _{\ast }:=\inf \left\{ \alpha \in (0,1)\text{ }|\text{ }\exists \beta
\in (0,1)\ /\ q\alpha +(1-q)\beta =p\text{ and }F_{X}^{-1}(\alpha)\geq
F_{Y}^{-1}(\beta)\right\}  \label{defal}
\end{equation}%
and $\beta _{\ast }=\frac{p-q\alpha _{\ast }}{1-q}.$ The proof consists in
verifying that $s_{p}$ can always be expressed as 
\begin{equation}
s_{p}=\max \left\{ F_X^{-1}(\alpha_*),F_Y^{-1}(\beta_*)\right\} .
\label{showthis}
\end{equation}%
From Table \ref{SUM}, it is clear that \eqref{showthis} is proved. Let us
now make the calculations case by case to prove Table \ref{SUM}.

\noindent \underline{Case 1: $F$ is continuous in $s_p$ and for all $z<s_p$, 
$F(z)<F(s_p)$}

\noindent In this case we always have that $s_p=F_X^{-1}(F(s_p)).$ Hence, we
only need to show that $\alpha _{\ast }=F(s_p)$ (i.e. $\beta _{\ast }=\frac{%
p-qF(s_p)}{1-q}$) and that $s_p=F_X^{-1}(\alpha _{\ast })\geq
F_Y^{-1}(\beta_*)$ as in this case (\ref{showthis}) will obviously hold.

Since $F_S^{-1}(p)=s_p$ then $F_S(s_p^-)=q F(s_p^-)+(1-q)G(s_p^-)\leqslant
p\leqslant F_S(s_p)=q F(s_p)+(1-q) G(s_p)$. Thus, by continuity of $F$, $q
F(s_p)+(1-q)G(s_p^-)\leqslant p\leqslant q F(s_p)+(1-q) G(s_p).$ Thus, 
\begin{equation}  \label{eqp}
G(s_p^-)\leqslant \frac{p-q F(s_p)}{1-q}\leqslant G(s_p)
\end{equation}

\noindent (1a): $G$ is continuous in $s_p$ and for all $z<s_p$, $%
G(z)<G(s_p). $ Then, $s_p=F_Y^{-1}(G(s_p)).$ It is also clear that for $%
\alpha <F(s_p) $ and thus $\beta >G(s_p),$ one has that $F_X^{-1}(%
\alpha)<F_Y^{-1}(\beta).$ Hence, as per definition of $\alpha _{\ast }$, one
has $\alpha _{\ast }=F(s_p),$ $\beta _{\ast }=G(s_p)$ and $%
s_p=F_X^{-1}(\alpha_{\ast })=F_Y^{-1}(\beta_*).$

\noindent (1b): $G$ is continuous in $s_{p}$ and there exists $z<s_{p},$ $%
G(z)=G(s_{p})$ $\ $(thus, $G$ is constant on the interval $(z,s_{p})$).
Then, $F_Y^{-1}(G(s_{p}))<s_{p}=F_X^{-1}(F(s_{p}).$ However, for $\alpha
<F(s_{p})$ and thus $\beta >G(s_{p}),$ one has that $F_X^{-1}(%
\alpha)<F_Y^{-1}(\beta).$ Hence, as per definition of $\alpha _{\ast },$ $%
\alpha _{\ast }=F(s_{p}),$ $\beta _{\ast }=G(s_{p})$ and $%
s_{p}=F_X^{-1}(\alpha_*)>F_Y^{-1}(\beta_*).$ Thus, $s_{p}=F_X^{-1}(%
\alpha_*)>F_Y^{-1}(\beta_*)$.

\noindent (1c): $G$ has a discontinuity in $s_{p}$ and for all $z<s_{p}$, $%
G(z)<G(s_{p}^{-}).$ From \eqref{eqp}, in this case, $F_Y^{-1}\left(\frac{%
p-qF(s_{p})}{1-q}\right)=s_{p}$. For $\alpha <F(s_{p})$ and thus $\beta >%
\frac{p-qF(s_{p})}{1-q},$ $F_X^{-1}(\alpha)<F_Y^{-1}(\beta).$ Hence, as per
definition of $\alpha _{\ast },$ $\alpha _{\ast }=F(s_{p}),$ $\beta _{\ast }=%
\frac{p-qF(s_{p})}{1-q}$ and $s_{p}=F_X^{-1}(\alpha_*)=F_Y^{-1}(\beta_*).$

\noindent (1d): $G$ has a discontinuity in $s_{p}$ and there exists $z<s_{p}$%
, $G(z)=G(s_{p}^{-})$ so that $G$ is constant on some interval $(r,s_{p})$
with $r<s_{p}$. From \eqref{eqp}, 
\begin{equation*}
F_Y^{-1}\left(\frac{p-qF(s_{p})}{1-q}\right)\leq s_{p}.
\end{equation*}%
If $\frac{p-qF(s_{p})}{1-q}>G(s_{p}^{-})$ (or equivalently, $%
F_{S}(s_{p}^{-})<p$)$,$ then $F_Y^{-1}\left(\frac{p-qF(s_{p})}{1-q}%
\right)=F_X^{-1}(F(s_{p})=s_{p}.$ Clearly, for $\alpha <F(s_{p})$ and thus $%
\beta >\frac{p-qF(s_{p})}{1-q},$ one has that $F_X^{-1}(\alpha)<F_Y^{-1}(%
\beta).$ Hence, as per definition of $\alpha _{\ast }$, one has $\alpha
_{\ast }=F(s_{p}),$ $\beta _{\ast }=\frac{p-qF(s_{p})}{1-q}$ and $%
s_{p}=F_X^{-1}(\alpha_*)=F_Y^{-1}(\beta_*).$ If $\frac{p-qF(s_{p})}{1-q}%
=G(s_{p}^{-})$ (or equivalently, $F_{S}(s_{p}^{-})=p$)$,$ then this implies
that $F_Y^{-1}\left(\frac{p-qF(s_{p})}{1-q}\right)<s_{p}.$ When $\alpha
<F(s_{p})$ thus $\beta >\frac{p-qF(s_{p})}{1-q}$ one has that $%
F_X^{-1}(\alpha)<F_Y^{-1}(\beta)\leq s_{p}.$ Hence, as per definition of $%
\alpha _{\ast }$, one has $\alpha _{\ast }=F(s_{p}),$ $\beta _{\ast }=\frac{%
p-qF(s_{p})}{1-q}$ and $s_{p}=F_X^{-1}(\alpha_{\ast })>F_Y^{-1}(\beta_*).$

\noindent \underline{Case 2: $F$ is continuous in $s_p$ and there is a $%
z<s_p,$ $F(z)=F(s_p)$ ($F(\cdot )$ is constant on $(z,s_p)$)}

\noindent (2a): this case can be obtained from (1b) by changing the role of $%
X $ and $Y$.

\noindent (2b): $G$ is continuous in $s_{p}$ and there exists $z<s_{p}$, $%
G(z)=G(s_{p}).$ Thus $G$ is constant on some interval $(r,s_{p})$ with $%
r<s_{p}$. Hence, $F_S^{-1}(p)\leqslant \min (z,z)<s_{p}$ which contradicts
the definition of $s_{p}=F_S^{-1}(p)$. The case (2b) is impossible.

\noindent (2c): $G$ is discontinuous in $s_{p}$ and for all $z<s_{p}$, $%
G(z)<G(s_{p}^{-}).$ From \eqref{eqp}, in this case, $F_Y^{-1}\left(\frac{%
p-qF(s_{p})}{1-q}\right)=s_{p}>F_X^{-1}(F(s_{p})$. However, for all $\alpha
>F(s_{p})$ and thus $\beta <\frac{p-qF(s_{p})}{1-q}$ it holds that $%
F_X^{-1}(\alpha)>F_Y^{-1}(\beta).$ Hence, as per definition of $\alpha
_{\ast },$ $\alpha _{\ast }=F(s_{p}),$ $\beta _{\ast }=\frac{p-qF(s_{p})}{1-q%
}$ and $s_{p}=F_Y^{-1}(\beta_*)>F_X^{-1}(\alpha_*).$\newline
\noindent (2d): $G$ is discontinuous in $s_{p}$ and there exists $z<s_{p}$, $%
G(z)=G(s_{p}^{-}).$ From \eqref{eqp}, $F_Y^{-1}\left(\frac{p-qF(s_{p})}{1-q}%
\right)\leq s_{p}.$ If $\frac{p-qF(s_{p})}{1-q}>G(s_{p}^{-})$ (or
equivalently, $F_{S}(s_{p}^{-})<p$)$,$ then $F_Y^{-1}\left(\frac{p-qF(s_{p})%
}{1-q}\right)=s_{p}>F_X^{-1}(F(s_{p}).$ For $\alpha >F(s_{p})$ and thus $%
\beta <\frac{p-qF(s_{p})}{1-q}$ one has that $F_X^{-1}(\alpha)>F_Y^{-1}(%
\beta).$ Hence, as per definition of $\alpha _{\ast }$, $\alpha _{\ast
}=F(s_{p}),$ $\beta _{\ast }=\frac{p-qF(s_{p})}{1-q}$ and $%
F_X^{-1}(\alpha_*)<F_Y^{-1}(\beta_*)=s_{p}.$ The case that $\frac{p-qF(s_{p})%
}{1-q}=G(s_{p}^{-})$ is excluded as it implies that $F_S^{-1}(p)<s_{p}$
should hold (similar to the case (2b)) which is a contradiction with the
definition of $s_{p}$.

\noindent \underline{Case 3: $F$ has a discontinuity in $s_{p}$ and for all $%
z<s_{p}$, $F(z)<F(s_{p}^{-})$}\newline
\noindent In this case, $s_{p}=F_X^{-1}(F(s_{p}).$ This situation is merely
identical to previous cases.

(3a): it is the same as (1c) by changing the role of $X$ and $Y$.

(3b): it is the same as (2d) by changing the role of $X$ and $Y$.

(3c): Observe that $F_X^{-1}(\alpha)=s_{p}$ for all $F(s_{p}^{-})\leq \alpha
\leq F(s_{p})$ and also that $F_Y^{-1}(\beta)=s_{p}$ for all $%
G(s_{p}^{-})\leq \beta \leq G(s_{p}).$

We also know that $F_{S}(s_{p}^{-})\leqslant p\leqslant F_{S}(s_{p})$ hence
there exists $F(s_{p}^{-})\leq \alpha _{1}\leq F(s_{p})$ and $%
G(s_{p}^{-})\leq \beta _{1}\leq G(s_{p})$ so that $q\alpha _{1}+(1-q)\beta
_{1}=p$ and $F_X^{-1}(\alpha_1)=F_Y^{-1}(\beta_1)=s_{p}.$ Therefore, $%
F_X^{-1}(\alpha_*)=F_Y^{-1}(\beta_*)=s_{p}.$

(3d): Observe that $F_X^{-1}(\alpha)=s_{p}$ for all $F(s_{p}^{-})\leq \alpha
\leq F(s_{p})$ and also that $F_Y^{-1}(\beta)=s_{p}$ for all $%
G(s_{p}^{-})<\beta \leq G(s_{p}).$

We also know that $F_{S}(s_{p}^{-})\leqslant p\leqslant F_{S}(s_{p})$ and
there are two possibilities:

In the case when $F_{S}(s_{p}^{-})<p$, then there exists $\alpha _{1}\in
(F(s_{p}^{-}),F(s_{p}))$ and $\beta _{1}\in (G(s_{p}^{-}),G(s_{p}))$ so that 
$q\alpha _{1}+(1-q)\beta _{1}=p$ and $F_X^{-1}(\alpha_1)=F_Y^{-1}(%
\beta_1)=s_{p}.$ Therefore, $F_X^{-1}(\alpha_*)=F_Y^{-1}(\beta_*)=s_{p}.$

In the case when $F_{S}(s_{p}^{-})=p,$ then $%
qF(s_{p}^{-})+(1-q)G(s_{p}^{-})=p$ and one has that $%
F_X^{-1}(F(s_{p}^{-}))>F_Y^{-1}(G(s_{p}^{-})_{\ast }),$ while for $\alpha
<F(s_{p}^{-})$ and $\beta >G(s_{p}^{-})$ one has that $F_X^{-1}(%
\alpha)<F_Y^{-1}(\beta).$ Hence, $\alpha _{\ast }=F(s_{p}^{-}),$ $%
G(s_{p}^{-})=\beta _{\ast }$ and $s_{p}=F_X^{-1}(\alpha_*)>F_Y^{-1}(\beta_*)$%
.

\noindent \underline{Case 4: $F$ has a discontinuity in $s_{p}$ and there
exists $z<s_{p}$, $F(z)=F(s_{p}^{-})$}\newline
By changing the role of $X$ and $Y$ we have that the case (4a) corresponds
to (1d), the case (4b) corresponds to (2d) and the case (4c) corresponds to
(3d). Finally the case of (4d) is treated as follows. In the case (4d), both 
$F$ and $G$ are discontinuous at $s_{p}$, and there exists $z_{1}$ and $%
z_{2} $ such that $F(z_{1})=F(s_{p})$ and $G(z_{2})=G(s_{p})$ so that $F$ is
constant on $(z_{1},s_{p})$ and $G$ is constant on $(z_{2},s_{p})$. Then $%
F_S^{-1}(p)\leqslant \min (z_{1},z_{2})<s_{p}$ which contradicts the
definition of $s_{p}=F_S^{-1}(p)$. This case is thus impossible.\hfill $\Box 
$

\bigskip 

It is clear that in many cases $F_{X}^{-1}(\alpha _{\ast })=F_{Y}^{-1}(\beta
_{\ast }).$ For example, it is sufficient that the distribution functions of 
$X$ and $Y$ are strictly increasing with unbounded support.

\bibliographystyle{econometrica}
\bibliography{biblio}

\end{document}